\RequirePackage{fix-cm}
\documentclass[twocolumn]{svjour3}          
\smartqed  
\usepackage{graphicx}
\begin{document}

\title{Video transrating in AVC and HEVC transcoding}


\author{Krzysztof Wegner \and Tomasz Grajek \and Jakub Stankowski \and 
Marek Doma\'nski}

\authorrunning{K. Wegner at al.} 

\institute{Chair of Multimedia Telecommunications\\and Microelectronics\\
			Pozna\'n University of Technology\\
			ul. Polanka 3, 60-965 Pozna\'n\\
            \email{(kwegner, tgrajek, jstankowski)@multimedia.edu.pl}
}

\maketitle

\begin{abstract}
HEVC (MPEG-H Part 2 and H.265) is a new coding technology which is expected to be deployed on the market along with new video services in the near future. HEVC is a successor of currently widely used AVC (MPEG-4 Part 10 and H.264). In this paper, the quality coding gains obtained for the Cascaded Pixel Domain Transcoder of AVC-coded material to HEVC standard are reported. Extensive experiments showed that transcoding with bitrate reduction allows the achievement of better rate-distortion performance than by compressing an original video sequence with the use of AVC at the same (reduced) bitrate.

\end{abstract}

\section{Introduction}
\label{intro}
Research on video transcoding techniques and algorithms is a very active topic, especially because of its significance for heterogeneous communication networks and a variety of user-end-systems \cite{Ref:1,Ref:2,Ref:3}. Every time when new coding technology is supposed to be deployed on market, those research become more intensive. Currently, a new video coding technology, HEVC - High Efficiency Video Coding (MPEG-H Part 2, H.265) \cite{Ref:4}, has been finalized by ISO/IEC MPEG and ITU-T VCEG. HEVC offers up to 50\% bitrate reduction in comparison to the commonly used AVC (MPEG-4 Part 10 and H.264) \cite{Ref:5}, while preserving the same subjective video quality \cite{Ref:6}. Due to its significantly higher compression efficiency, it is expected that new video systems (e.g., 4k video, video streaming) that exploit HEVC will be deployed in the near future. Due to a huge amount of legacy video material, new transcoding scenarios involving the HEVC technique will probably attract a lot of attention.

Most of the works currently being published focus on reducing the quality losses caused by transcoding or/and complexity reduction \cite{Ref:7,Ref:8,Ref:9,Ref:10,Ref:11,Ref:12} in comparison to the so called Cascaded Pixel Domain Transcoder (CPDT) \cite{Ref:3}. The CPDT is the most straightforward transcoder configuration that consists of a full decoder followed by an encoder. In other words we get knowledge about have different approaches to transcoding utilizing information about motion vectors \cite{Ref:9,Ref:11,Ref:13}, image partitioning \cite{Ref:9,Ref:10}, selected modes \cite{Ref:10} can speedup this process and also influence the coding efficiency. However, always in relation to the CPDT. But, what might be expected from CPDT approach? It is obvious that transcoding introduces inevitable and irreversible quality loss. However, HEVC provides more efficient data representation than AVC. Therefore, it would be very interesting to know which effect (quality loss caused by re-quantization, or more efficient data representation) is stronger.

Another aspect of great practical importance in transcoding is transrating (transcoding that leads to bitrate change) \cite{Ref:7,Ref:9,Ref:12,Ref:13}. There are many situations when we would like to have a smaller video bitstream for example in order to fit to communication channel or smaller file size to simply save space on the hard drive. Therefore, when original video is unavailable (most of the cases) we can transcode our compressed video to the lower bitrate. But what can be achieved in terms of coding efficiency, when during transrating we additionally change the compression standard (in our case from AVC to HEVC)?
 
In the paper, we will clearly show that transcoding from AVC to HEVC can bring not only no quality losses, but surprisingly, even objective quality gains in terms of PSNR. Such conclusions are drawn from extensive experiments concerning CPDT from AVC to HEVC. The authors are not aware of any reference to CPDT rate-distortion performance, such as presented in this paper.

\section{Methodology}
\label{sec:1}
\begin{figure}
\includegraphics[width=\linewidth]{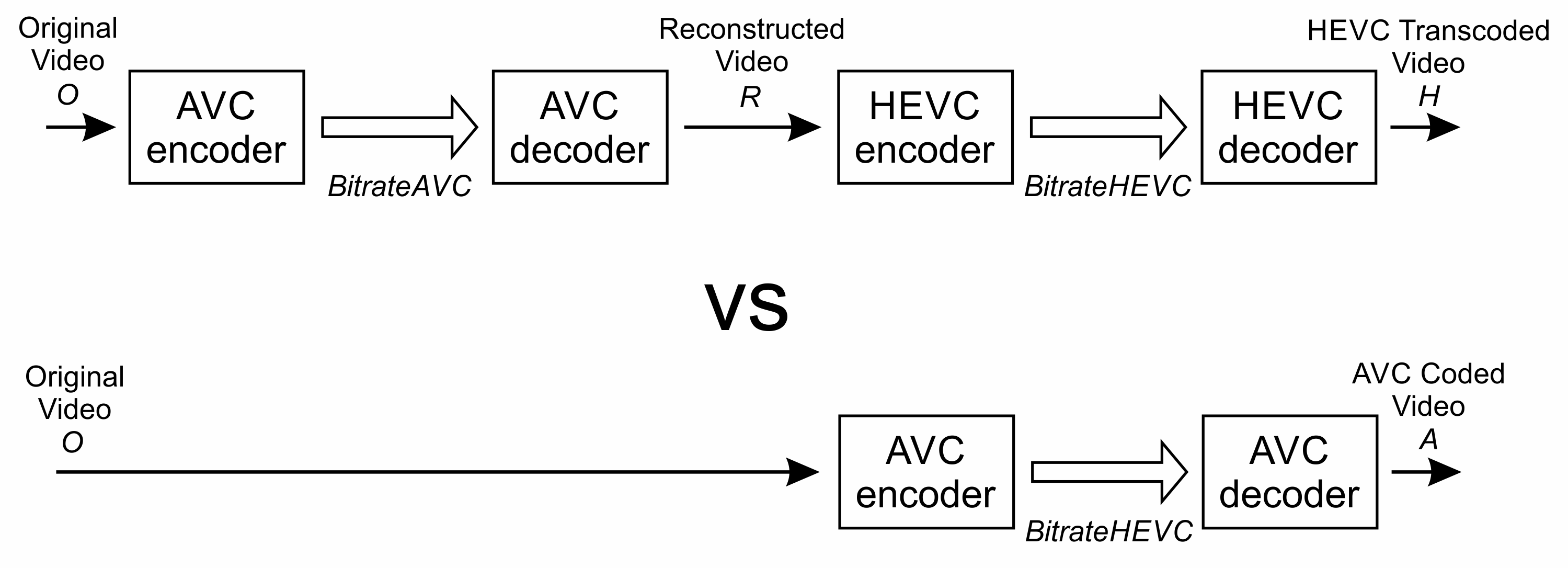}
\caption{Methodology of the experiments.}
\label{fig:1}       
\end{figure}

In order to evaluate the performance of CPDT from AVC to HEVC, a number of test video sequences were encoded using AVC for a wide range of bitrates (bitrate controlled by $QP$). Then, each AVC-encoded bitstream were transcoded to HEVC, again for a wide range of bitrates (controlled by $QP$ value). The bitrates were gathered before and after transcoding along with a corresponding quality of the decoded material in terms of the luminance PSNR metric (always in relation to the original uncompressed sequence). This allows a comparison of rate-distortion curves after transcoding with those obtained for AVC. Scheme of performed experiments has been shown in Fig. \ref{fig:1}.

\begin{figure}
\includegraphics[width=\linewidth]{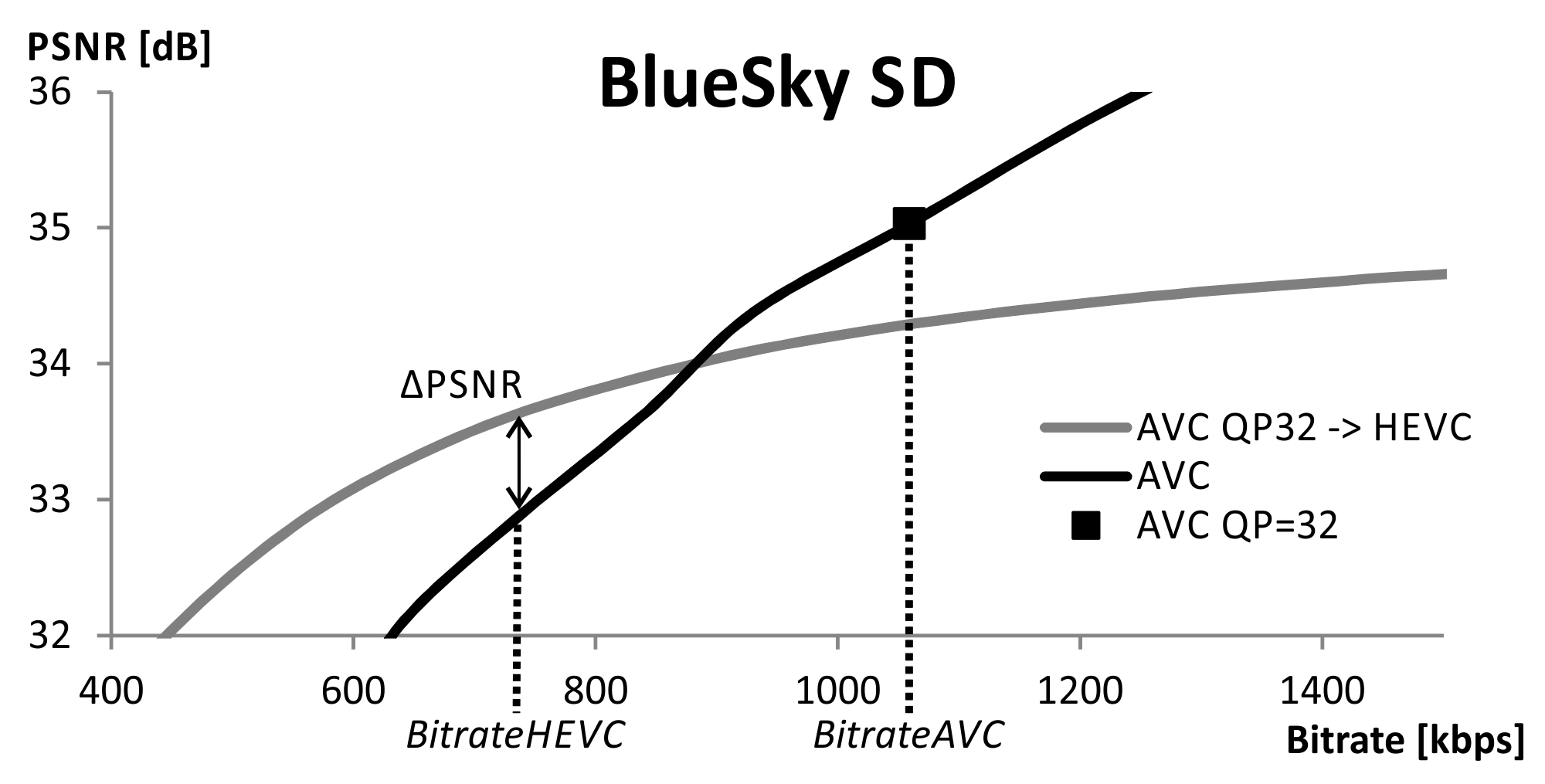}
\caption{Example of $\Delta$PSNR calculations for CPDT from AVC to HEVC. 'AVC QP32$\rightarrow$HEVC' describes the rate-distortion curve achieved for the transcoding of  material encoded with the use of AVC with $QP=32$ (starting point) to HEVC.}
\label{fig:2}       
\end{figure}

For the transcoded material, the PSNR difference ($\Delta$PSNR) were calculated. The $\Delta$PSNR is defined as the difference between the quality of the material transcoded with the use of HEVC and the quality of the original material that could potentially be encoded with the use of AVC at the same bitrate as the HEVC-transcoded one (see also Fig. \ref{fig:2}):
\begin{equation}
\Delta PSNR= 10\cdot log(\frac{N\cdot 255^{2}}{\sum (H-O)^{2}})-10\cdot log(\frac{N\cdot 255^{2}}{\sum (A-O)^{2}})
\end{equation}

\section{Experiments}
\label{sec:2}
All experiments according to presented methodology were conducted on a wide set of video sequences recommended by ISO/IEC MPEG as video test material for video compression technique evaluation during AVC development. The video sequences test set used covers a wide range of content characteristics in legacy video material. In total, we used 19 sequences: 7 -- HD (1920x1080): $Bluesky$, $Pedestrian$, $Riverbed$, $Rushhour$, $Station2$, $Sunflower$, $Tractor$ and 12 -- SD (704x576): $Bluesky$, $City$, $Crew$, $Harbour$, $Ice$, $Pedestrian$, $Riverbed$, $Rushhour$, $Soccer$, $Station2$, \\$Sunflower$ and $Tractor$. For the production of AVC-encoded material, the H.264/AVC reference software JM 18.4 \cite{Ref:14} with all possible $QP$ values from the range $10\div50$ was used. Each bitstream, after being decoded, was again encoded with an HEVC reference software version HM 15.0 \cite{Ref:15}, again with all possible $QP$ values from the range $10\div50$. This resulted in $12\cdot41\cdot41=20172$ transcodings for SD sequences and $7\cdot41\cdot41=11767$ transcodings for HD sequences. Both encoders were configured according to a sets of conditions, recommended by ISO/IEC MPEG, which are broadly used by scientific community for comparison of compression techniques (i.e. for AVC \cite{Ref:16} and for HEVC \cite{Ref:17}). Table \ref{tab:1} presents the essential configuration parameters used for AVC and HEVC encoders.

\begin{table*}
\caption{Essential configuration parameters used for AVC and HEVC encoders.}
\label{tab:1}       
\begin{tabular}{lll}
\hline\noalign{\smallskip}
Parameter & AVC & HEVC\\
\noalign{\smallskip}\hline\noalign{\smallskip}
Profile & Main for SD sequences & Main for SD and HD sequences\\
 &  High for HD sequences & \\
GOP & IBBPBBP… & IBBB…\\
 & GOP size = 16 & GOP size = 16\\
Hierarchical GOP & No & Yes\\
No. of ref. frames & 5 & 4\\
Rate-Distortion Optimization & On & On\\
Search range for Motion Estimation & ±16 for SD / 
±32 for HD & ±64\\
Entropy coding & CABAC & CABAC\\
\noalign{\smallskip}\hline
\end{tabular}
\end{table*}

\section{Results}
\label{sec:3}
\begin{figure}
\includegraphics[width=\linewidth]{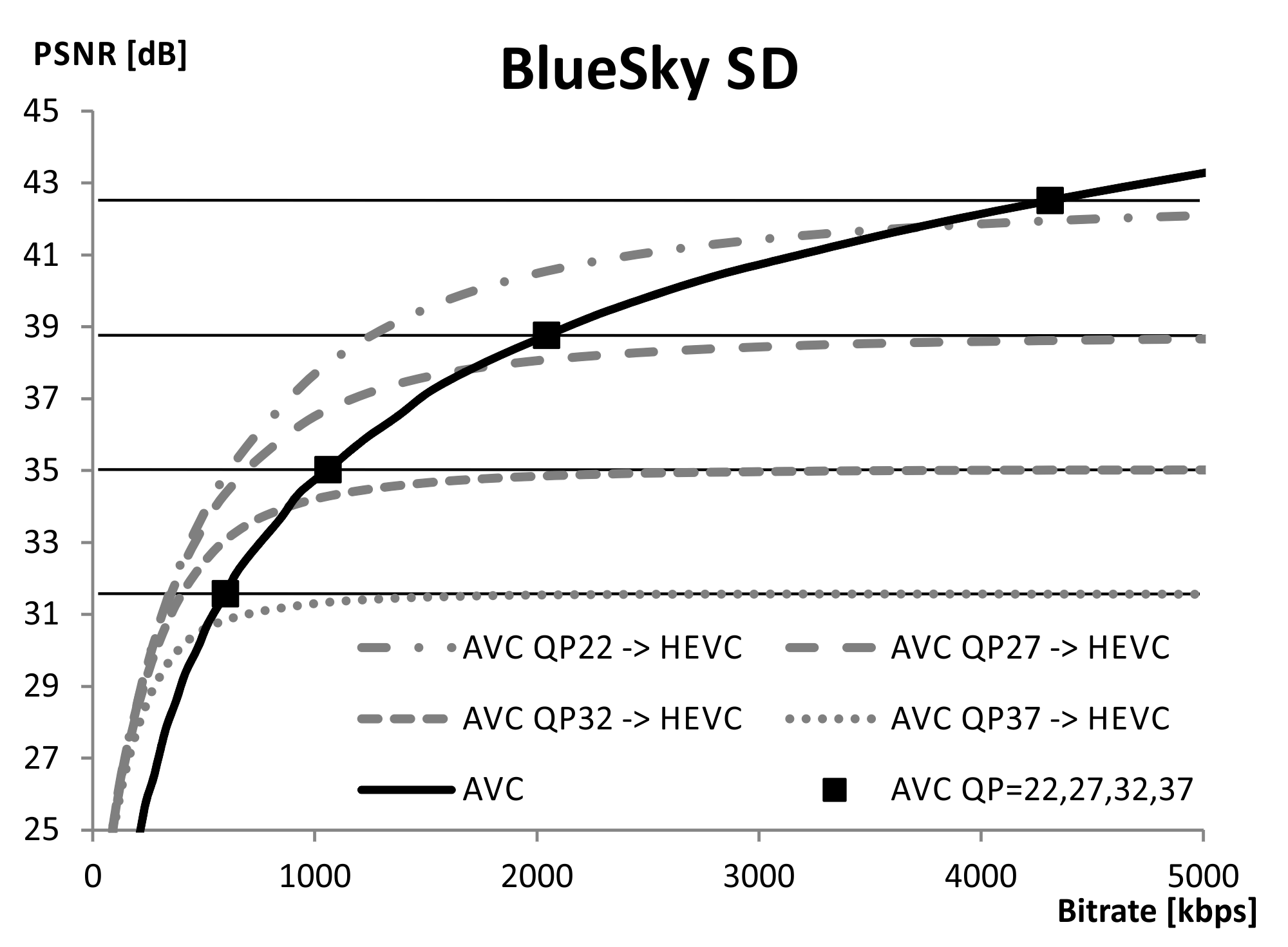}
\caption{Exemplary results for CPDT from AVC to HEVC for BlueSky SD sequence. 'AVC QP22$\rightarrow$HEVC' describes the rate-distortion curve achieved for the transcoding of material encoded with the use of AVC with $QP=22$ (starting point) to HEVC.}
\label{fig:3}       
\end{figure}

Exemplary results for the BlueSky sequence (SD resolution) has been shown in Fig. \ref{fig:3}. The black line is the rate-distortion curve for AVC-encoded material. The black square points represent exemplary starting points used for transcoding (i.e., AVC-encoded material at different $QP$ values). Additionally, the horizontal black dotted lines show the quality (PSNR for luminance in relation to the original sequence) of each starting point. Obviously, none of the HEVC-transcoded material created on the basis of this starting point can exceed this line in terms of quality. Based on the starting points, the grey lines were created, representing HEVC-transcoded material, one line per each starting point. For the transcoded material, the PSNR difference ($\Delta$PSNR) was calculated for luminance component (Y).

The $\Delta$PSNR with respect to the bitrate ratio between HEVC-transcoded material ($BitrateHEVC$) and its AVC-coded starting point ($BitrateAVC$) for SD and HD sequences, have been shown in Fig. \ref{fig:4} and \ref{fig:5} respectively. Additionally, the black line shows the quality difference ($\Delta$PSNR) versus bitrate reduction (caused by transcoding) averaged over all sequences.

\section{Conclusions}
\label{sec:4}
\begin{figure}
\includegraphics[width=\linewidth]{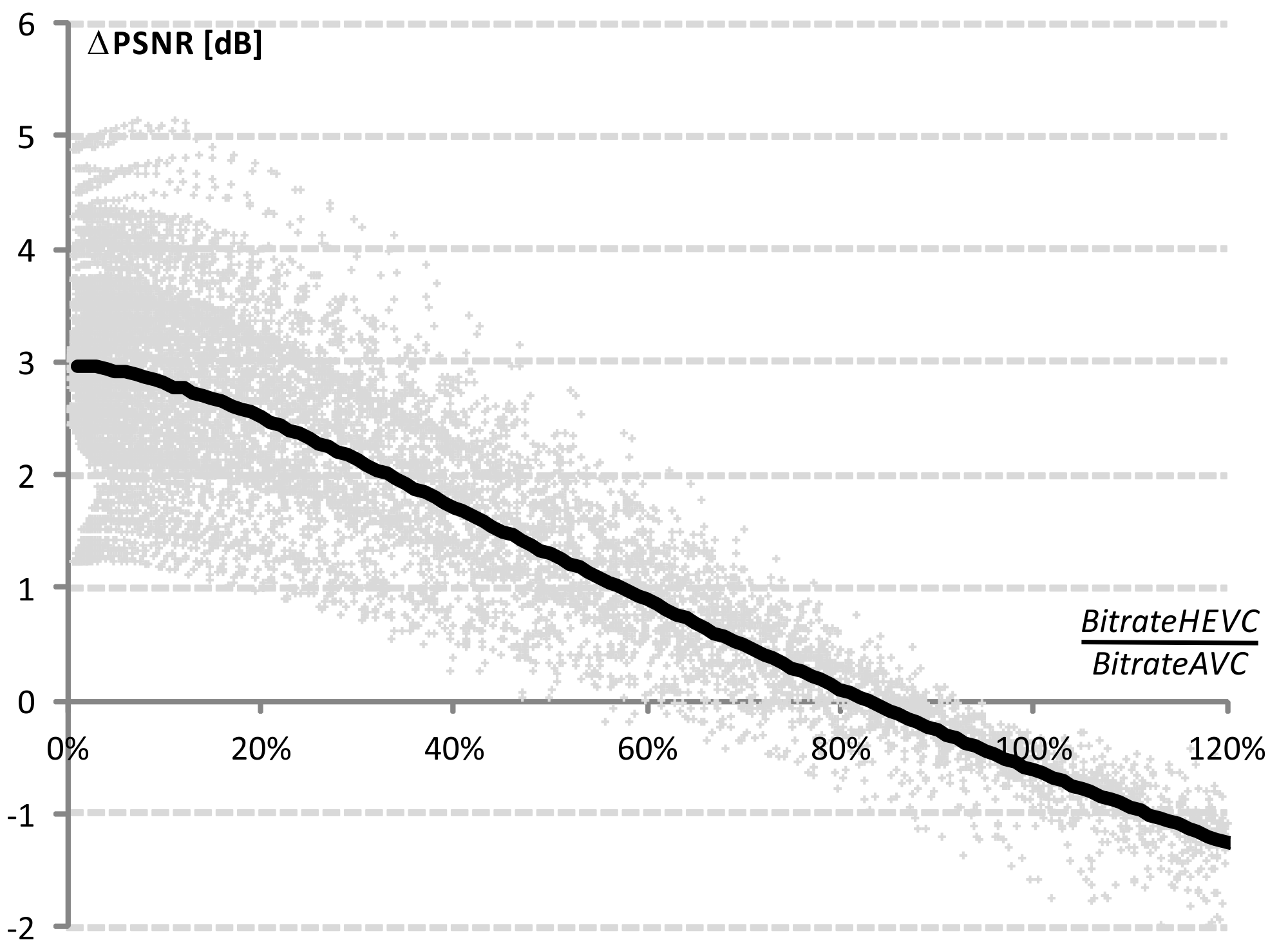}
\caption{$\Delta$PSNR with respect to bitstream reduction after transcoding for SD resolution sequences. The black line indicates the quality difference averaged over all SD sequences and all $QP$ values used.}
\label{fig:4}       
\end{figure}

\begin{figure}
\includegraphics[width=\linewidth]{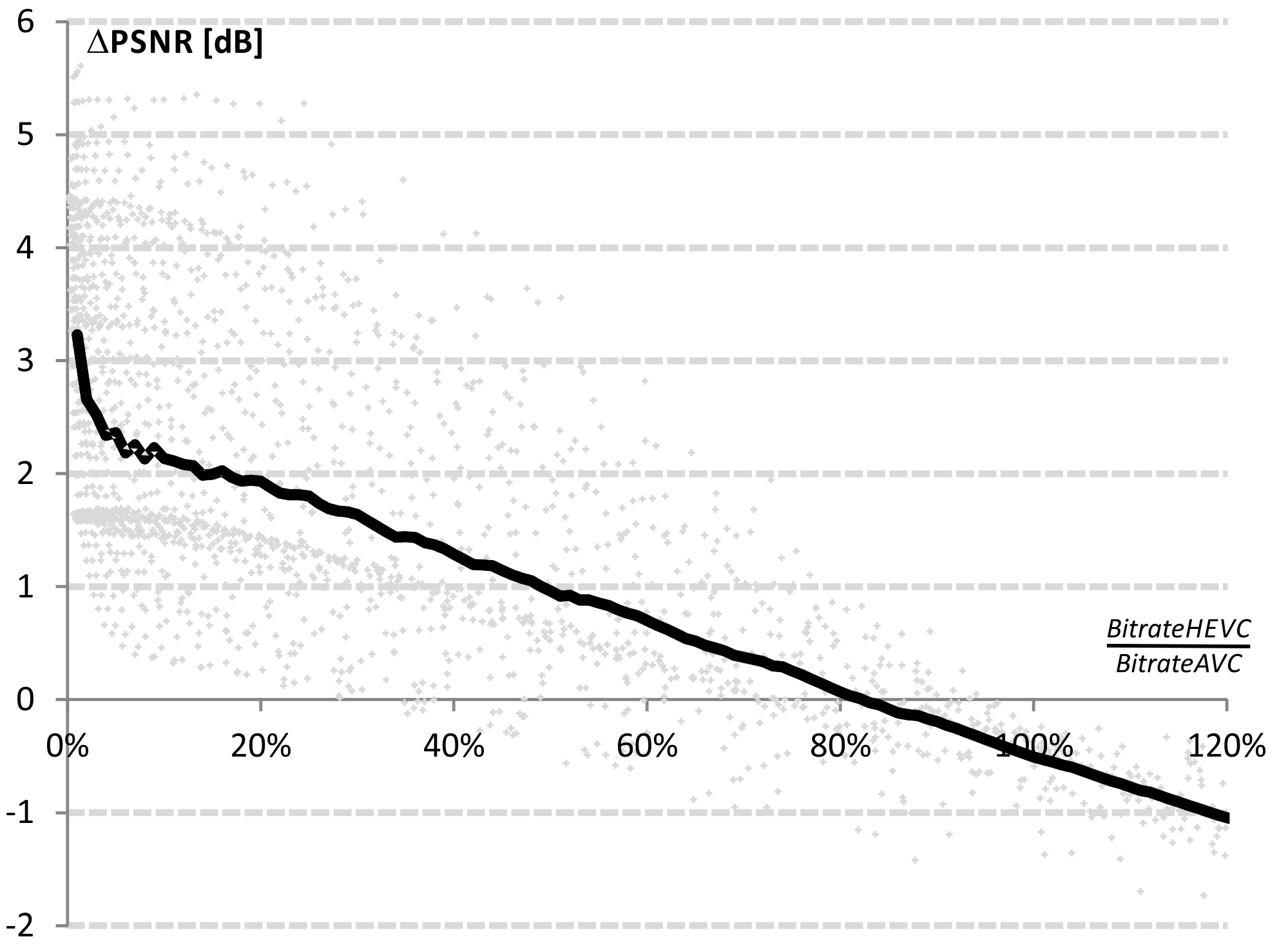}
\caption{$\Delta$PSNR with respect to bitstream reduction after transcoding for HD resolution sequences. The black line indicates the quality difference averaged over all HD sequences and all $QP$ values used.}
\label{fig:5}       
\end{figure}

Surprisingly, as shown in Fig. \ref{fig:4} and \ref{fig:5}, the transcoding of AVC-encoded material to HEVC with the bitrate reduction of 20\% or more gives, on average, objective quality gains with respect to AVC rate-distortion characteristics. It means that a video sequence, once transcoded to HEVC, is represented more efficiently than with the use of AVC. Therefore, despite the quality degradation caused by re-quantization (transcoding), it is possible to achieve a better rate-distortion performance compared to compressing the original video sequence with the use of AVC. Such relation was not observed for any other transcoders between previous compression techniques.

Of course, transcoding from AVC to HEVC at the same bitrate causes an average quality loss of 0.5dB for both SD and HD sequences (see Fig. \ref{fig:4} and \ref{fig:5}). Therefore, transcoding with preserving the same bitrate obviously causes a reduction in signal representation efficiency.

To conclude, transcoding from AVC to HEVC along with bitrate reduction might be considered as a very promising approach in practical applications.

\begin{acknowledgements}
Research project was supported by The National Centre for Research and Development, POLAND, Grant no. LIDER/023/541/L-4/12/NCBR/2013.
\end{acknowledgements}

\end{document}